\documentclass[aps,prb,preprint,showpacs,floatfix]{revtex4}
\usepackage{graphicx}

\begin{document}
\title{Energy Levels in Spheroidal Quantum Dot}

\author
{Anjana Bagga$^{\star}$, P. K. Chattopadhyay$^{\dag}$ and Subhasis Ghosh$^{\star} $}

\affiliation{$ ^{\star}$School of Physical Sciences, Jawaharlal Nehru University, New Delhi 110067\\
             $ ^{\dag}$Department of Physics, Maharshi Dayanand University, Rohtak}

\begin{abstract}
The effect of non-sphericity of the  quantum dot on the eigenvalues and eigenfunctions has been investigated for the case of both the finite and infinite barrier. The ground and excited state energies have been calculated for prolate and oblate spheroids as a function of eccentricity of the spheroid. The analytic wavefunctions giving the admixture of higher angular momentum states  have been obtained as a function of eccentricity.
\end{abstract}


\maketitle

The theoretical and experimental investigation of quantum dots(QDs) continues to attract a lot of attention. Recent experiments have demonstrated that the optical properties of QDs can be changed by controlling their shape\cite{XGP00}$^-$\cite{LSL01}. Non-spherical QDs are found to give rise to radiation which is linearly polarized\cite{JH01} in contrast to that from a spherical dot. In considering the formation of dark and bright excitons in QDs, Efros et al\cite{ALE96} have shown the importance of the non-sphericity of the QDs for the study of the stokes shift. The large width of optical transitions in QDs has been attributed to the distribution of both shapes and sizes of QDs. The lowering of spatial symmetry in QDs by changing the shape can be used to uncover an interesting class of physical laws in different quantum systems\cite{CWJB97}. The effect of the departure from sphericity on the eigenfunction and eigenvalues of the dots have been studied theoretically by several authors\cite{ALE93}$^-$\cite{VAB02}. Efros and Rodina\cite{ALE96} have calculated the electronic energy levels in ellipsoidal QDs using perturbation theory over spherical dots. Cantele et al\cite{GC00}$^-$\cite{DN01} have studied the energy levels of particles confined to an ellipsoidal quantum dot, solving the Schrodinger equation numerically and also by variational methods. They have studied the optical transition matrix elements and the oscillator strength in different size and shape of QDs.
Fonoberov et al\cite{VAB02} have calculated electron-hole energy spectrum and Coulomb potential energy in tetrahedral QD's  using the finite difference method.   In this work we calculate the energy levels of particles confined in a ellipsoidal QDs using spheroidal harmonics. The eigenvalues and eigenfunctions are obtained as expansions in terms of a parameter related to the eccentricity of the ellipse generating the ellipsoid. One obtains analytic wave functions revealing the extent of mixing of higher values of orbital angular momentum as the departure from sphericity increases. This method is readily extended to QDs where the barrier at the surface of the ellipsoid is not infinite. This is of great practical importance since the QDs are usually either grown on substrates using different epitaxial methods or embedded in a matrix of large band gap material such as glass, polymers, liquids etc. In all these cases, the electrons reaching the surface will encounter a finite not an infinite barrier. The optical properties of $QD$s are expected to depend on the height of the barrier which will be taken as a parameter.

{\sl Eigenvalues and Eigenfunctions of a Particle in an Ellipsoid}: 
A suitable co-ordinate system to investigate the motion of particles confined to the interior of a prolate (oblates) ellipsoid is the prolate (oblate) spheroidal co-ordinates.
We shall first consider the prolate case, and the oblate case later. The prolate spheroidal co-ordinates $\xi,\eta,\theta$ are related to the Cartesian co-ordinates x,y,z  by the relations:
\begin{equation}
\label{h1}
\begin{array}{lll}
x &=& {{\it f} \over 2} [(1-\eta^2)(\xi^2-1)]^{1/2} \cos \theta,\\
y &=& {{\it f} \over 2} [(1-\eta^2)(\xi^2-1)]^{1/2} \sin \theta,~~~
z = {{\it f} \over 2} \eta \xi
\end{array}
\end{equation}
with $~~~-1 \leq \eta \leq 1,~~$ $1 \leq \xi \leq \infty,~~$ $0 \leq \theta \leq 2\pi$.
In the prolate spheroidal system the surface $\xi=$constant $>1$ is an ellipsoid with major axis of length ${\it f}\xi$ and minor axis of length ${\it f}(\xi^2-1)^{1/2}$.  f is the confocal distance and $\xi = { 1 \over e}$, where {\it e} is the eccentricity of the ellipsoid.
The surface $|\eta|=$constant $<1$ is a hyperboloid of two sheets with an asymptotic cone whose generating line passes through the origin and is inclined at the angle $\vartheta = \cos^{-1} \eta$ to the $z-$axis. The surface $\varphi=$constant is a plane through the $z-$axis forming the angle $\varphi$ with the $x,z-$plane.

The ellipsoid is assumed to be generated by the rotation around $z-$axis of an ellipse having either the major or the minor axis along $z$. The semi-axis along $z$ will be denoted by `a' and the transverse axis as `b'. Hence the parameter $\chi = a/b$ is a measure of departure from sphericity. $\chi > 1$ is for a prolate ellipsoid and $\chi < 1$ is for  oblate ellipsod. 
The Schrodinger equation of the particle inside the ellipsoid
\begin{equation}
\label{h2}
- {\hbar^2 \over 2m^*} \nabla^2 \psi = E\psi
\end{equation}
where $m^*$ is the effective mass, 
The solution of Eqn.{\ref {h2}} can be given by 
$~~\psi_{lm} = S_{lm} (c,\eta) R_{lm} (c,\xi) e^{im\theta} ~~$
 which gives the following ordinary differential equations for $S_{lm} (c,\eta)$ and $R_{lm} (c,\xi)$
\begin{equation}
\label{h5}
{d \over d\eta} \Bigg[ (1-\eta^2) {d \over d\eta} S_{lm} (c,\eta)\Bigg] + \Bigg[ \lambda^{(c)}_{lm} - c^2 \eta^2 - {m^2 \over 1-\eta^2} \Bigg] S_{lm} (c,\eta) = 0
\end{equation}
and
\begin{equation}
\label{h6}
{d \over d\xi} \Bigg[ (\xi^2-1) {d \over d\xi} R_{lm} (c,\xi)\Bigg] - \Bigg[ \lambda^{(c)}_{lm} - c^2 \xi^2 + {m^2 \over \xi^2-1} \Bigg] R_{lm} (c,\xi) = 0
\end{equation}
where $c={1 \over 2} kf$, $k=\hbar\sqrt{2m^*} $. In order that $\psi_{lm}$ be single valued $m$ must be an integer. For $c=0$ the Eqn.(\ref{h5})) reduces to the associated Legendre equation. Therefore it follows that $\lambda_{lm} (0) = l(l+1),~~~~~ l \geq m$ and $S_{lm} (0,\eta) = P^m_l (\eta)$
where $P^m_l (\eta)$ are the associated Legendre polynomials. $S_{lm} (c,\eta)$ for $c\neq 0$ can be expanded in an infinite series of the form\cite{CF57}
\begin{equation}
\label{h9}
S_{lm} (c,\eta) = {\sum\limits^{\infty}_{r=0,1}}^\prime d^{lm}_r (c) P^m_{m+r} (\eta)
\end{equation}
The prime over the summation indicates that the summation is over only even values of $r$ when $(l-m)$ is even and only odd values of $r$ when $(l-m)$ is odd. The requirement that the wavefunction is finite at $\eta = \pm 1$ confines the $\eta$ - dependence of the wavefunction to that of the angle functions of the first kind. Substitution of (\ref{h9}) in Eqn.(\ref{h5}) leads to recurrence relations involving the coefficients $d^{lm}_{r-2} (c)$, $d^{lm}_r (c)$ and $d^{lm}_{r+2} (c)$.  For small values of $c^2$ both $\lambda_{mn} (c)$ and $d^{lm}_{l-m+\nu}/d^{lm}_{l-m}$, where $r=n-m+\nu$, can be expanded in power series of the form:
\begin{equation}
\label{h10}
\begin{array}{cc}
\lambda_{lm} (c) = \sum\limits_{k=0} q^{lm}_{2k} c^{2k},   &~~
 {d^{lm}_{l-m+2} \over d^{lm}_{l-m}} = \sum\limits_{k=1} p_{2k} c^{2k}
\end{array}
\end{equation}
The procedure to calculate the coefficients $q^{lm}_{2k}$ and $p_{2k}$ is outlined in Appendix.

The radial functions $R_{lm} (c,\xi)$ are obtained from the functions $S_{lm} (c,\xi)$ using  the relation\cite{CF57}
\begin{equation}
\label{h12}
R_{lm} (c,\xi) = \int\limits^b_a e^{ic\eta \xi} (\xi^2-1)^{m/2} (1-\eta^2)^{m/2} S_{lm} (c,\eta) d\eta,
\end{equation}
These are the radial functions of the first kind which are appropriate for the particles confined to spheroid problem and are given by
\begin{equation}
\label{h13}
R_{lm} (c,\xi) = {\rho_{lm} (\xi^2-1)^{m/2} \over (c\xi)^m} \sum\limits^{\infty 1}_{r=0,1} d^{lm}_r (c) i^{r} {(2m+r)! \over r!} j_{m+r} (c\xi)
\end{equation}
where $\rho_{lm}$ is a normalization factor and $j_{m+r} (c\xi)$ are spherical Bessel function.

The co-ordinate $\xi$ is related to the eccentricity $e$ of the ellipse of the revolution by $\xi=1/e$. The radial functions $R_{lm} (c,\xi)$ must satisfy the boundary condition that at the surface of the ellipsoid defined by $\xi=\xi_1=1/e$, the wave function vanishes. This leads to the equation
\begin{equation}
\label{h14}
\sum\limits^{\infty}_{r=0,1} d^{lm}_r (c) i^r {(2m+r)! \over r!} j_{m+r} (c\xi_1) = 0
\end{equation}
With the known expressions of $d^{lm}_r (c)$, Eqn.(\ref{h14}) can be solved to obtain $c$. Since $c=\frac{1}{2} kf$, the energy eigenvalues are given by
\begin{equation}
\label{h15}
E_{nlm} = {\hbar^2 k^2_{nlm} \over 2m^*} = {\hbar^2 \over 2m^*} \Bigg( {2c_{nlm} \over f} \Bigg)^2
\end{equation}
the subscripts $l$ and $m$ of $c$ indicates that the solutions will depend on $l$ and $m$. The subscript $n$ implies that, in general, there will be more than one solution corresponding to different zeros of Eqn.\ref{h14}.

 The corresponding equations for the oblate spheroid can be obtained from those of the prolate spheriod by making the transformation $\xi \rightarrow +i\xi, c \rightarrow -ic$\cite{CF57} in Eqn.{\ref{h5}} and {\ref{h6}}.
The eigenvalues of energy of a particle in an oblate spheriod are determined by Eqn.\ref{h15} where c is now a solution of
\begin{equation}
\label{h19}
\sum\limits^{\infty}_{r=0,1} d^{lm}_r (-ic) i^r {(2m+r)! \over r!} j_{m+r} (c\xi_1) = 0
\end{equation}

The energy eigenvalues Eqn.{\ref{h15}} for the prolate and oblate spheriod case can be put in the form
\begin{equation}
E_{nlm} = \Bigg( {\hbar^2 \over 2m^* \alpha^2} \Bigg) \epsilon_{nlm}(\chi)   
\end{equation}
where $\alpha=a$ in case of prolate and $\alpha=b$ for oblate. $\epsilon_{nlm} (\chi)$ are the energies in units of $(\hbar^2/2m^* \alpha^2)$ which depend on the ellipsoid aspect ratio $\chi = a/b$ but not on $a$ or $b$ separately. The energy levels as functions of $\chi$ for both finite and infinite barriers are shown in Fig.1. For prolate case $(\chi > 1)$ the major axis along $z$ is increased keeping the minor axis in the transverse direction fixed at the spherical value. In oblate case $(\chi < 1)$ the minor axis is along $z$ and is held fixed at the spherical value while the major axis in the transverse direction is increased. The volume of the spheroid generated by rotation of the ellipse around the $z-$axis, $\frac{4}{3} \pi ab^2$, can be written as $\frac{4}{3} \pi R^3 (a/R)$ for the prolate and $\frac{4}{3} \pi R^3 (b/R)^2$ for the oblate case where $R$ is the radius of the sphere corresponding to the case $a=b=R$. (We follow the convention that the semi-axis of the ellipsoid along $z$, major or minor, is denoted by `a').

It is observed that the degeneracy w.r.t. $m$ present in the spherical case $(\chi = 1)$ is removed as one moves away from sphericity. In the prolate case as $\chi$ increases and in the oblate case as $\chi$ decreases the energies decrease which can be easily understood as caused by the deconfining effect due to increase in volume. For a particular $l$ and $m$, energy increases as the barrier height increases, the energy for the infinite barrier being the highest. This is again due to the fact that for infinite barrier height the particles are confined to a smaller volume. The smaller the height of the barrier at the surface, the bigger is the probability of finding the particle outside in a bigger volume and hence the smaller is the energy.

It is observed that the energy of a state for a prolate with a given value of $(\frac{a}{b})$ is higher compared to that for an oblate spheroid with the same value for $(\frac{b}{a})$. This is easily explained by the fact that the volume for the oblate case is greater than that for the prolate with the corresponding value of $(\frac{a}{b})$. For a given $\chi$ the $(l,m)=(1,0)$ state has higher energy than the (1,1) state for oblate spheroid. The situation is just the reverse in the prolate case. For the (1,1) state the motion is mostly confined to the $xy$ plane whereas for the (1,0) state the motion is along the $z-$direction. Since in the oblate case the dimensions along the $xy$ direction are greater than that in the $z-$direction, it is expected that the (1,1) energy will be less than the (1,0) energy in the oblate case due to deconfinig effects. For similar reasons the (1,0) state is lower than the (1,1) state for the prolate spheriod.

In Fig.2 the difference between the energies of the states for the spherical and spheroidal QDs($E_{sphere} - E_{spheroid}$) is plotted as a function of eccentricity $e$ for the ground state of the prolate spheriod in CdSe and GaN materials. Keeping the axis in the transverse direction fixed at $16$\mbox{\AA}, the axis `a' along the $z-$axis is increased thereby increasing the eccentricity. The increase of the quantity $(E_{sphere} - E_{spheroid})$ with $e$ implies that the energies of the states decrease with increasing eccentricity. This is equivalent to the observation earlier in Fig.1 that the energies decrease with $ \chi $. The inset of Fig.2 shows the variation of the energies as a function of $e$ when the volume is held fixed. In this case $E_{spheriod} > E_{sphere}$ and increases with $e$ at a faster rate. As the dimensions along the $z-$direction is increased (giving rise to a decrease in kinetic energy), the dimensions along the transverse directions must decrease giving rise to an increase in kinetic energy. The net effect is an increase in energy. The corresponding results for  oblate spheriod case are shown in Fig.3 and the inset for constant volume. Keeping the axis along $z$ fixed at $16$\mbox{\AA} as the transverse axes are increased, the energies decrease with $e$. In this case $E_{spheriod} < E_{sphere}$ and results are just the opposite when the volume is kept fixed.

In Tables:1 and 2 are given the coefficients of $j_{o}(c\xi)$ and $j_{2}(c\xi)$ in the radial
wavefunction {Eqn.\ref{h13} of the ground state  (l,m)=(0,0) as a function of eccentricity in the oblate and prolate case respectively, when the volume is held fixed. For both the infinite and finite barrier, the admixture of $j_{2}(c\xi)$ increases as the eccentricity
increases with a corresponding decrease in the $j_{o}(c\xi)$  component, but for the finite barrier case the rate of increase is much less as compared to that when the barrier is infinite.

In conclusion, the energy eigenvalues for both the prolate and oblate ellipsoids have been calculated as a function of eccentricity for both the finite and infinite barrier case.
The analytic expressions for the wavefunctions have been obtained. It has been shown that for both the infinite and finite barrier, the probability of states with higher values of angular momentum increases as the eccentricity increases, the rate of increase being smaller for the finite case.  In the context of recent experimental findings\cite{XGP00}$^-$\cite{LSL01}, the results presented here illuminate the  important role of shape in addition to the size of QDs  on   the  possibility of tailoring their electronic and optical properties.


\newpage
\begin{center}
\section*{Table I.}
{\footnotesize
\begin{tabular}{|c|l|l|l|l|l|}  \hline
             & \multicolumn{2}{c|}{Infinite barrier} & \multicolumn{2}{c|}{Finite barrier}\\ \cline{2-5}
Eccentricity & \multicolumn{2}{c|}{coefficient of}   & 
 \multicolumn{2}{c|}{coefficient of}  \\ \cline{2-5}
      (e)       & \multicolumn{1}{c|}{$j_{o}(c\xi)$} & \multicolumn{1}{c|}{$j_{2}(c\xi)$} & 
 \multicolumn{1}{c|}{{$j_{o}(c\xi)$}} & \multicolumn{1}{c|}{{$j_{2}(c\xi)$}} \\ \hline
     .1      & .999     &  $1.22~X~10^{-4}$ &  .999 &  $1.59~X~10^{-5}$   \\ \hline
     .3      & .989     &  $1.09~X~10^{-2}$ &  .998 &  $1.30~X~10^{-3}$   \\ \hline
     .5      & .903     &  .097             &  .988 &  $1.15~X~10^{-2}$   \\ \hline
     .6      & .788     &  .211             &  .974 &  $2.55~X~10^{-2}$   \\ \hline
     .7      & .610     &  .384             &  .942 &  $5.78~X~10^{-2}$   \\ \hline
\end{tabular}
}
\end{center}
\begin{center}
\section*{Table II.}
{\footnotesize
\begin{tabular}{|c|l|l|l|l|l|}  \hline
             & \multicolumn{2}{c|}{Infinite barrier} & \multicolumn{2}{c|}{Finite barrier}\\ \cline{2-5}
Eccentricity & \multicolumn{2}{c|}{coefficient of}   & 
 \multicolumn{2}{c|}{coefficient of}  \\ \cline{2-5}
     (e)        & \multicolumn{1}{c|}{$j_{o}(c\xi)$} & \multicolumn{1}{c|}{$j_{2}(c\xi)$} & 
 \multicolumn{1}{c|}{{$j_{o}(c\xi)$}} & \multicolumn{1}{c|}{{$j_{2}(c\xi)$}} \\ \hline
     .1      & .999     &  $1.21~X~10^{-4}$ &  .999 &  $1.59~X~10^{-5}$   \\ \hline
     .3      & .985     &  $1.03~X~10^{-2}$ &  .998 &  $1.31~X~10^{-3}$   \\ \hline
     .5      & .917     &  .083             &  .987 &  $1.22~X~10^{-2}$   \\ \hline
     .6      & .828     &  .171             &  .974 &  $2.56~X~10^{-2}$   \\ \hline
     .7      & .706     &  .289             &  .946 &  $5.36~X~10^{-2}$   \\ \hline
\end{tabular}
}
\end{center}

\newpage
\section*{Table Captions}
\noindent

\vspace*{.5cm}
\noindent
Table I :
The coefficients of $j_{o}(c\xi)$ and $j_{2}(c\xi)$ in the radial wavefunction $Eqn.{\ref{h13}}$ of the ground state as a function of eccentricity for the oblate case.

\vspace*{.5cm}
\noindent
Table II :
The coefficients of $j_{o}(c\xi)$ and $j_{2}(c\xi)$ in the radial wavefunction $Eq.{\ref{h13}}$ of the ground state as a function of eccentricity for the prolate case.

\newpage
\section*{Figure Captions}
\noindent

\vspace*{.5cm}
\noindent
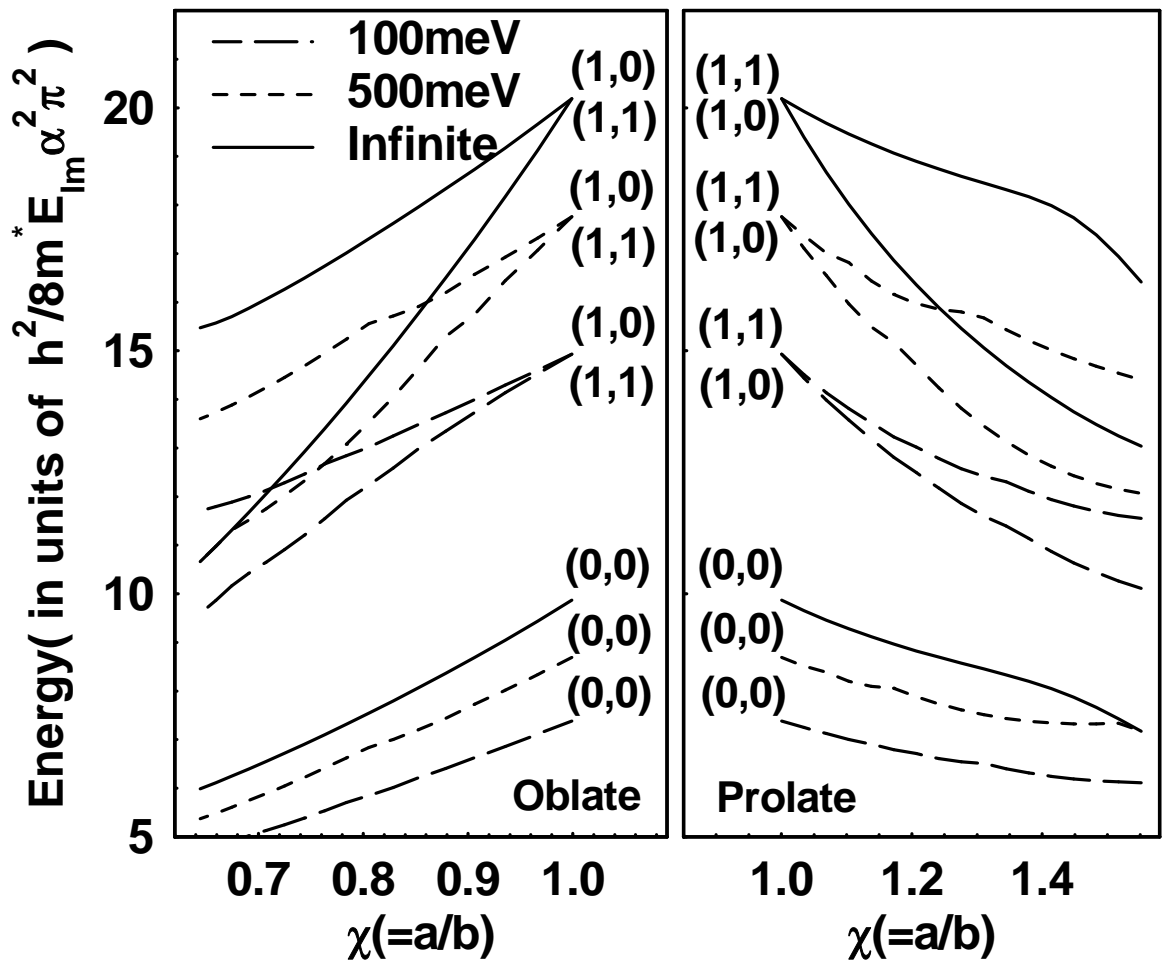
Fig.1:
Eigenvalues as a function of $\chi=\frac{a}{b}$ where a is the intercept along the z-axis 
and b along the transverse axis. $\alpha$=a and b for the prolate and oblate,  respectively.
      
\vspace*{.5cm}
\noindent
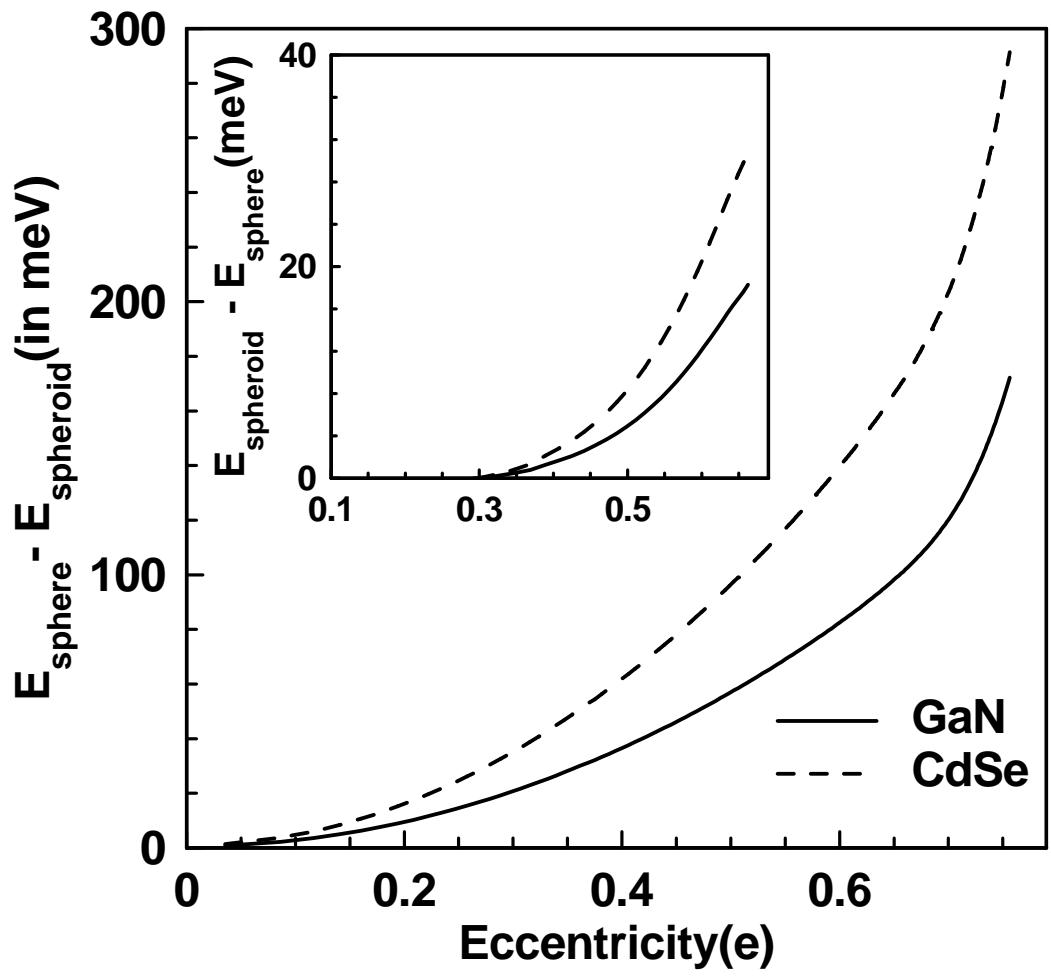
Fig.2: The difference between the eigenvalues for the spherical and spheroidal QDs  as a function of eccentricity  for the ground state of prolate spheroid in CdSe and GaN materials. The minor axis is fixed and major axis is increased thereby increasing the eccentricity. The inset gives the same for the constant volume. 

\vspace*{.5cm}
\noindent
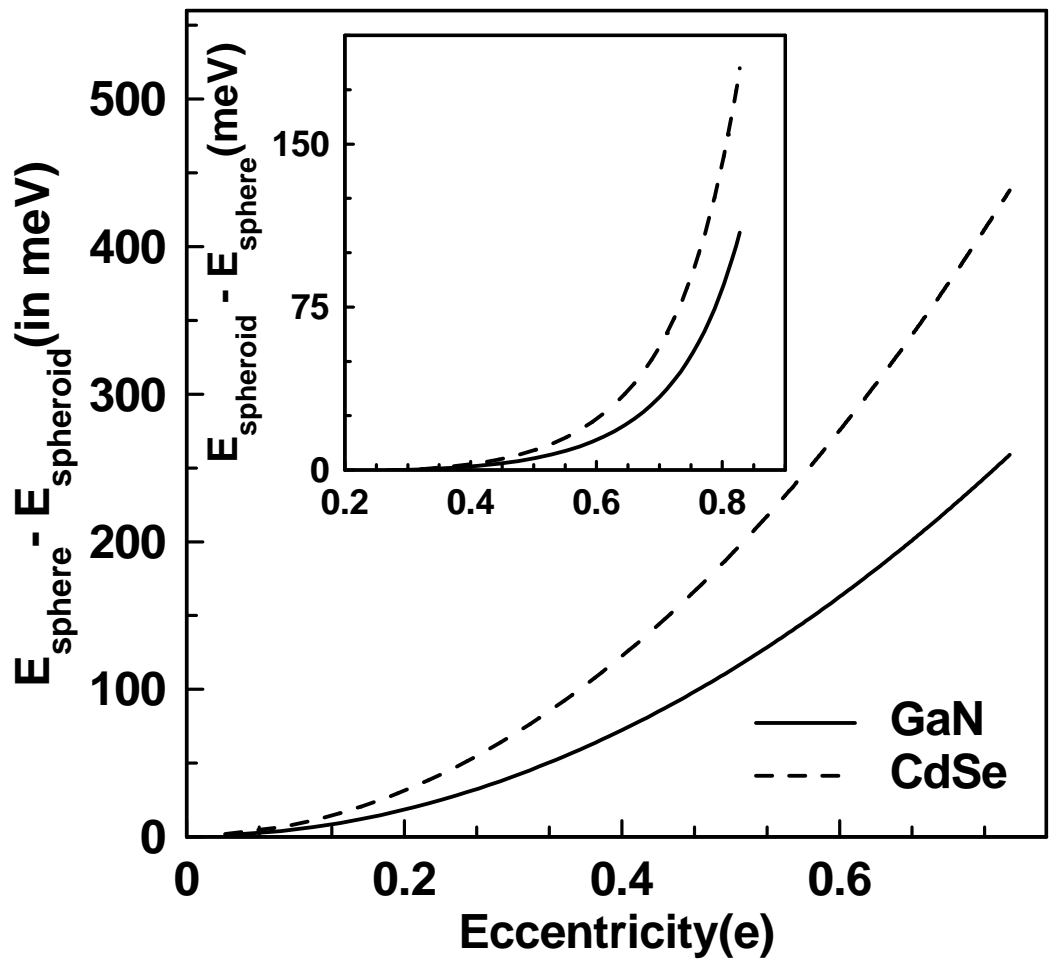
Fig.3: The difference between the eigenvalues for the spherical and spheroidal QDs as a function of eccentricity  for the ground state of oblate spheroid for CdSe and GaN materials. The minor axis is fixed and major axis is increased thereby increasing the eccentricity. The inset gives the same for the constant volume.

\newpage
\section*{Appendix}
The recurrence relation involving the coefficients $d^{lm}_r (c)$ obtained by substituting
Eqn.(\ref{h9}) in Eqn.(\ref{h5}) can be given by
\begin{equation}
\label{Ap1}
\begin{array}{l}
 {(l+m+2+\nu)(l+m+1+\nu) \over (2l+3+2\nu)(2l+5+2\nu)} c^2 d^{lm}_{l-m+\nu+2}(c)+\nonumber\\ 
 \Bigg[ (l+\nu)(l+\nu+1) - \lambda_{lm}(c) + {2(l+\nu)(l+\nu+1)-2m^2-1 \over (2l+2\nu-1)(2l+2\nu+3)} c^2 \Bigg] d^{lm}_{l-m+\nu} \nonumber\\
 + {(l-m+\nu)(l-m-1+\nu) \over (2l+2\nu-3)(2l+2\nu-1)} c^2 d^{lm}_{l-m+\nu-2} = 0,~~~~ (r \geq 0)
\end{array}
\end{equation}
where $ \nu = 0, \pm 2, \pm 4,.....$
Putting ${\nu=0}$ in Eqn.{\ref{Ap1}} one obtains the expansion of $\lambda_{lm}(c)$ in powers of $c^2 :$%
\begin{equation}
\label{Ap2}
\begin{array}{l}
\lambda_{lm}(c)  = l(l+1) + {2(l)(l+1)-2m^2-1 \over (2l-1)(2l+3)} c^2 \nonumber\\ 
 +  {(l+m+2)(l+m+1) \over (2l+3)(2l+5)} c^2 {d^{lm}_{l-m+2} \over d^{lm}_{l-m}} 
 + {(l-m)(l-m-1) \over (2l-3)(2l-1)} c^2 {d^{lm}_{l-m-2} \over d^{lm}_{l-m}}
\end{array}
\end{equation}
In Eqn.{\ref{Ap2} both ${d^{lm}_{l-m+2} \over d^{lm}_{l-m}}$ and ${d^{lm}_{l-m-2} \over d^{lm}_{l-m}}$ are given by power series in $c^2 :$
\begin{eqnarray}
\label{Ap3}
{d^{lm}_{l-m+2} \over d^{lm}_{l-m}} = p_2c^2+p_4c^4+... and~
{d^{lm}_{l-m-2} \over d^{lm}_{l-m}} = p_2^{'} c^2+p_4^{'}c^4+... 
\end{eqnarray}
Putting $\nu=2$ in Eqn.{\ref{Ap1} one obtains 
\begin{equation}
\label{Ap4}
\begin{array}{l}
 \Bigg[ (l+2)(l+3)-\lambda_{lm}(c)+{2(l+2)(l+1)-2m^2-1 \over (2l+3)(2l+7)} c^2 \Bigg]  {d^{lm}_{l-m+2} \over d^{lm}_{l-m}} \\
= - {(l-m+2)(l-m+1) \over (2l+1)(2l+3)} c^2 - {(l+m+4)(l+m+3) \over (2l+7)(2l+9)} c^2 {d^{lm}_{l-m+4} \over d^{lm}_{l-m}}
\end{array}
\end{equation}
Substituting Eqn.{\ref{Ap2}, {\ref{Ap3} in Eqn.{\ref{Ap4}, noting that ${d^{lm}_{l-m+4} \over d^{lm}_{l-m}}=O(c^4)$ and comparing the coefficients of $c^2$ and $c^4$ from both sides one obtains
$p_2$ and $p_4$. Similarly putting $\nu=-2$ in Eqn.{\ref{Ap1}} and following a similar procedure, $p_2^{'}$ and $p_4^{'}$ can be determined. $p_6,~p_8..$ and $p_6^{'},~p_8^{'}..$
in Eqn.{\ref{Ap3}} are obtained by putting $\nu=\pm4,\pm6$ etc. and repeating the procedure.

\end{document}